\documentclass[preprint,amssymb,amsmath]{article}

\usepackage{graphicx}
\usepackage{pifont}
\usepackage{textcomp}
\graphicspath{{fig/}}
\usepackage{dcolumn}
\usepackage{bm}
\usepackage{float}
\usepackage{amssymb}
\usepackage{amsmath}
\usepackage{color}
\begin{document}

\title{Thermal spin pumping mediated by magnon    \\
in the semiclassical regime  }

\author{Kouki Nakata,    \\    \   \\
 Yukawa Institute for Theoretical Physics, 
Kyoto University,  \\
Kitashirakawa Oiwake-Cho, Kyoto 606-8502, Japan,  \\
nakata@yukawa.kyoto-u.ac.jp}

\maketitle

\begin{abstract}
We microscopically analyze thermal spin pumping mediated by magnons, 
at the interface between a ferromagnetic insulator and a non-magnetic metal, in the semiclassical regime. 
The generation of a spin current is discussed by calculating the thermal spin transfer torque, which breaks the spin conservation law for
conduction electrons and operates the coherent magnon state.
Inhomogeneous  thermal fluctuations between conduction electrons and magnons 
induce a net spin current, which is pumped into the adjacent  non-magnetic metal. 
The pumped spin current is proportional to the  temperature difference.
When the effective temperature of magnons is lower than that of conduction electrons,
localized spins lose spin angular momentum by emitting magnons and conduction electrons flip from down to up 
by absorbing all the emitted momentum, and vice versa. 
Magnons at the zero mode cannot contribute to thermal spin pumping 
because they are eliminated  by the spin-flip condition. 
Consequently thermal spin pumping does not cost any kinds of applied magnetic fields.
We have discussed the distinction from the theory proposed by Xiao et al. [Phys. Rev. B, ${\mathbf{81}}$ (2010) 214418],
Adachi et al. [Phys. Rev. B, ${\mathbf{83}}$ (2011) 094410],
and Bender et al. [arXiv:1111.2382]. \footnote{
{\textit{Supplement is available at this URL}};
http://dl.dropbox.com/u/5407955/SupplementTSP.pdf
}
\end{abstract}

\maketitle

\section{Introduction}
\label{sec:intro}

Recently spintronics  has developed a new branch of physics
called spin caloritronics  \cite{caloritronics, Bauerspincaloritronics},
which combines thermoelectrics with spintronics. 
Spin caloritronics  has been attracting a special interest
because of potential applications
to green information and communication technologies \cite{mod}.
The central theme is the utilization of thermal fluctuations
as well as spin degrees of freedom 
in order to induce a (pure) spin current. 
Thus establishing methods for the generation of a spin current
by using thermal difference, 
without any kinds of applied magnetic fields, is a significant issue.

In the previous work  \cite{QSP}, we have studied quantum spin pumping mediated by magnons under a time-dependent transverse magnetic field 
at the interface between a ferromagnetic insulator and a non-magnetic metal. 
There the ferromagnet act as a source of spin angular momentum; magnon battery named after the spin battery \cite{battery}.
The applied time-dependent transverse magnetic field acts as a quantum fluctuation
to induce a pumped net spin current under a thermal equilibrium condition.
Spin angular momentum is exchanged between conduction electrons and localized spins via magnons
accompanying the exchange interaction at the interface.
The interface is defined as an effective area where the Fermi gas (conduction electrons)  and 
the Bose gas (magnons) coexist to interact;
the width of the interface is  supposed to be of the order of the lattice constant  \cite{width}.
In addition, the pumped net spin current has a resonance structure as a function
of the angular frequency of the applied transverse field, 
which is useful to enhance the spin pumping effect induced by quantum fluctuations.
Here
it should be stressed that 
magnons accompanying the exchange interaction  
cannot contribute to spin pumping without quantum  fluctuations.
That is, quantum fluctuations (i.e. time-dependent transverse magnetic fields)
are essential to quantum spin pumping mediated by magnons.

\begin{figure}[h]
\begin{center}
\includegraphics[width=7cm,clip]{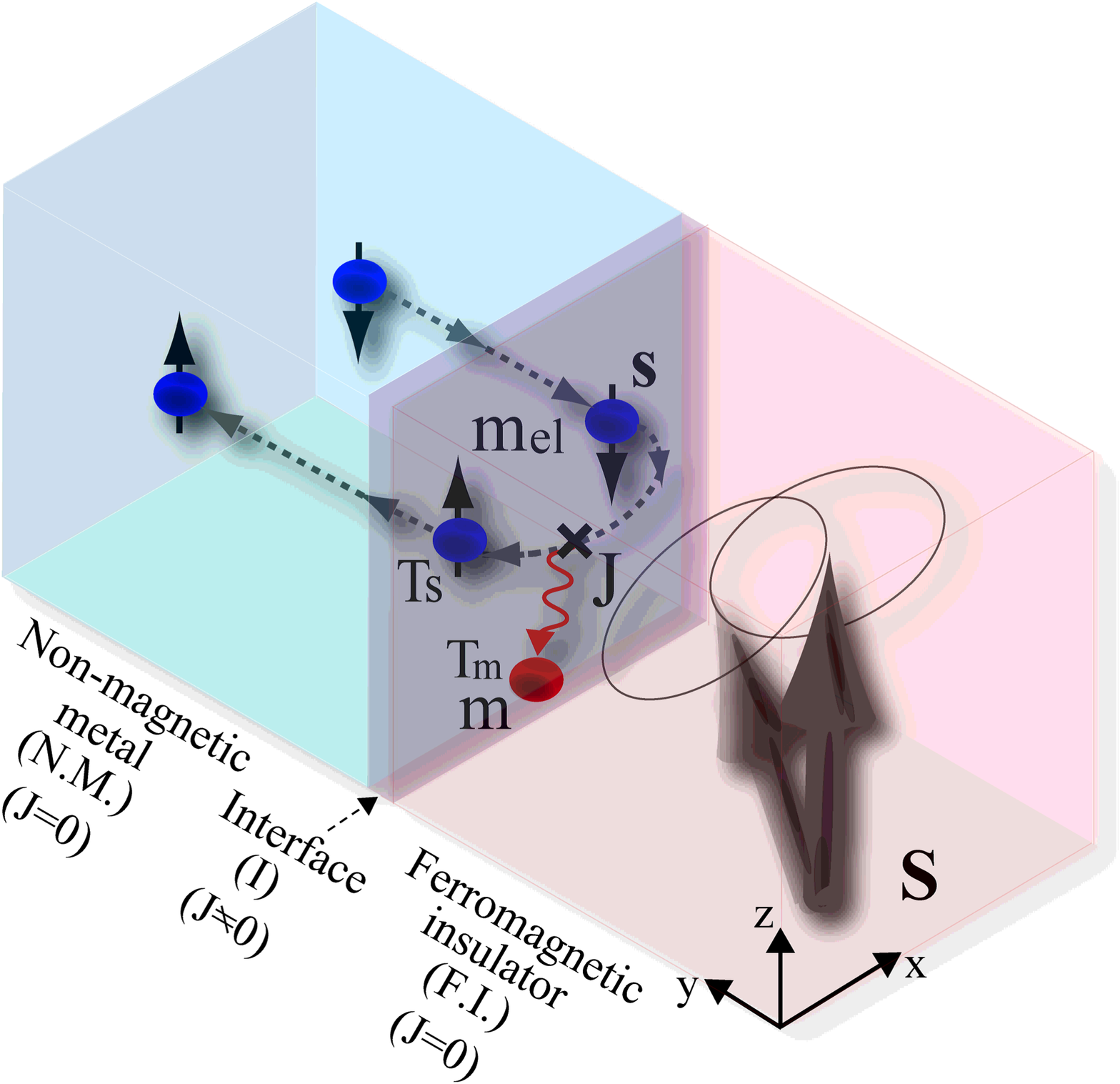}
\caption{
A schematic picture of thermal spin pumping mediated by magnons.
Spheres  represent magnons and
those with arrows are conduction electrons.
When the effective temperature of magnons ($T_{\rm{m}}$) is lower than that of conduction electrons  ($T_{\rm{s}}$),
localized spins lose spin angular momentum by emitting a magnon 
and conduction electrons flip from down to up by absorbing the momentum,
and vice versa. 
The interface is defined as an effective area where 
the Fermi gas (conduction electrons)  and 
the Bose gas (magnons) coexist to interact; $J\not=0$.
In addition,
conduction electrons  cannot  enter the ferromagnet,
which is an insulator. 
 \label{fig:pumping} }
\end{center}
\end{figure}

In this paper, we microscopically propose an alternative mechanism for the generation of the spin current 
without any kinds of applied magnetic fields (i.e. quantum fluctuations); thermal spin pumping  \cite{uchidathermal}.
Inhomogeneous   thermal fluctuations, i.e. the temperature difference,
between conduction electrons and magnons 
induce a net spin current, which is pumped into the adjacent  non-magnetic metal.
This  method can be viewed as an alternative way for the local spin injection.

We assume the local equilibrium condition \cite{zubarev};
since the relaxation times in the localized spins (i.e. magnons)  and conduction electrons subsystems are 
much shorter than the lattice relaxation time   \cite{twotemperature,twotemperatureprl, twotemperaturesov}, 
the reservoirs become thermalized internally before they equilibrate with each other. 
Therefore  we may assume that  during the relaxation process,
conduction electrons and magnons  can be described 
by their effective local temperatures; $T_{\rm{s}}$ and $T_{\rm{m}}$  \cite{uchidathermal,adachi}.
According to Xiao et al.  \cite{xiao},
the condition (i.e. temperature difference) can be generated by a temperature bias applied over the ferromagnetic film.

The theoretical setup \cite{adachi} is  almost  the same with our previous work  \cite{QSP} except the point that 
applied magnetic fields are not essential;
in particular, transverse magnetic fields are absent.
We consider   a ferromagnetic insulator and  non-magnetic   metal   junction
shown in Fig. \ref{fig:pumping}
where conduction electrons couple with localized spins  $  {\mathbf{S}}({\mathbf{x}},t) $, ${\mathbf{x}}=(x, y, z) \in {\mathbf{R}}^3$, 
at the interface;
\begin{equation}
 {\cal{H}}_{\rm{ex}}=  - 2Ja_0^3  {\int_{{\mathbf{x}}\in \rm{(interface)}}} d {\mathbf{x}}   {\ }  {\mathbf {S}}({\mathbf{x}},t) \cdot   {\mathbf {s}}({\mathbf{x}},t).
\label{eqn:e1}
\end{equation}
The exchange coupling constant reads  $ 2J $,  and the lattice constant of the ferromagnet is  $a_0$.
 In this paper, we take $\hbar =1$ for convenience. 
The magnitude of the interaction is supposed to be constant and 
we adopt  the continuous limit also in the present study.
Conduction electron spin variables are represented as  
\begin{eqnarray}
 {{s}^{j}}  = \sum_{\eta ,\zeta  = \uparrow , \downarrow} [c^{\dagger }_{\eta} (\sigma  ^j)_{\eta \zeta} c_{\zeta}]/2    \\
      \         \         \              \equiv  (c^{\dagger }  \sigma ^j c)/2,
 \end{eqnarray}                  
where $  {  {\sigma }^j} $ are the $ 2\times 2 $  Pauli matrices;
$ [ \sigma  ^j , \sigma  ^k  ] = 2i\epsilon _{jkl} \sigma  ^l   $, ($ j, k, l = x,y,z$).
Operators $c^{\dagger }/c $ are   creation/annihilation operators for conduction electrons, 
which satisfy the (fermionic) anticommutation relation;
$ \{c_{\eta  }({\mathbf{x}}, t), c_{\zeta }^{\dagger }({\mathbf{x}}', t) \}= \delta _{\eta , \zeta } \delta ({\mathbf{x}}-{\mathbf{x'}})  $.

We focus on the dynamics at the interface where spin angular momentum is exchanged 
between conduction electrons and the ferromagnet.
We suppose the uniform magnetization and
thus  localized spin degrees of freedom   can be mapped into magnon ones 
via the Holstein-Primakoff transformation;
\begin{eqnarray}
   S^{+} ({\mathbf{x}},t)   \equiv S^{x} ({\mathbf{x}},t)+iS^{y} ({\mathbf{x}},t)  \\
\      \           \                     \       \       \      \        \       \       \      \  =  \sqrt{2\tilde S} a({\mathbf{x}},t) + {\cal{O}}({\tilde S}^{-1/2}),    
\end{eqnarray}                 
\begin{eqnarray}
   S^{-}({\mathbf{x}},t) \equiv S^{x} ({\mathbf{x}},t)+iS^{y} ({\mathbf{x}},t)   \\
\      \           \                     \       \       \      \        \       \       \      \ =   \sqrt{2\tilde S}  a^\dagger ({\mathbf{x}},t)    + {\cal{O}}({\tilde S}^{-1/2}),    
   \end{eqnarray}    
\begin{eqnarray}
   S^z({\mathbf{x}},t) = \tilde S-a^\dagger ({\mathbf{x}},t)  a   ({\mathbf{x}},t),
      \end{eqnarray}
$\tilde S\equiv  S/{a_0^3}$,                    
where operators $a^{\dagger }/a $ are magnon creation/annihilation operators
satisfying the (bosonic) commutation relation; 
$ [a(\mathbf{x}, t), a^{\dagger }(\mathbf{x'}, t) ]= \delta (\mathbf{x}-\mathbf{x'}) $.
Up to the $ {\cal{O}}(S)$ terms,
localized spins  reduce to a free boson system.
Consequently in the quadratic dispersion (i.e. long wavelength) approximation, 
the  localized spin   with the applied magnetic field along the quantization axis (z-axis) is described
by the Hamiltonian $  {\cal{H}}_{\rm{mag}} $;
\begin{eqnarray}
  {\cal{H}}_{\rm{mag}}  =  \int_{{\mathbf{x}}\in \rm{(interface)}} d   {\mathbf{x}}  {\ }  {a^{\dagger }(\mathbf{x},t)} 
                                             \Big(-\frac{ { \mathbf{\nabla }}^2 }{2m} + B \Big)  
                                             {a(\mathbf{x},t)},
\label{eqn:e2}
\end{eqnarray}
and the Hamiltonian,  ${\cal{H}}_{\rm{ex}} (\equiv     {\cal{H}}_{\rm{ex}} ^{S}+{\cal{H}}_{\rm{ex}} ^{\prime})$, can be rewritten as
\begin{eqnarray}
 {\cal{H}}_{\rm{ex}} ^{S}   =   -JS  {\int_{{\mathbf{x}}\in \rm{(interface)}}} d    {\mathbf{x}}  
                                                         {\ }   c^{\dagger } ({\mathbf{x}},t)  \sigma^z  c  ({\mathbf{x}},t), 
\label{eqn:e3}                                                                                                                                                               
\end{eqnarray}
\begin{eqnarray}
{\cal{H}}_{\rm{ex}} ^{\prime} = -Ja_0^3  \sqrt{\frac{\tilde S}{2}} {\int_{{\mathbf{x}}\in \rm{(interface)}}} d   {\mathbf{x}}  
                                      [ a^{\dagger }({\mathbf{x}},t)  c^{\dagger }  ({\mathbf{x}},t) \sigma ^{+} c({\mathbf{x}},t )  
                                 +  a({\mathbf{x}},t)  c^{\dagger } ({\mathbf{x}},t)  \sigma ^{-} c({\mathbf{x}},t )].
\label{eqn:e4}
\end{eqnarray}
The variable 
 $ m $ represents the effective mass of a magnon.
We have denoted a constant applied magnetic field along the quantization axis as  $B$,
which includes $g$-factor and Bohr magneton.
Let us mention that 
though we formulate the thermal spin pumping theory with $B$ for generalization, in this paper
we finally take $B=0$ in sec. \ref{sec:kel} and discuss the thermal spin pumping effect in sec. \ref{sec:pumping}.

The  total Hamiltonian of the system (interface), ${\cal{H}}$,  is given as 
\begin{eqnarray}
{\cal{H}}  =   { \cal{H}}_{\rm{mag}} +  { \cal{H}}_{\rm{ex}}^{\prime} +  { \cal{H}}_{\rm{el}},   \  \rm{where}
\end{eqnarray}
\begin{eqnarray}
{\cal{H}}_{\rm{el}} = \int_{{\mathbf{x}}\in \rm{(interface)}} d   {\mathbf{x}}   {\ }  c^{\dagger } ({\mathbf{x}},t) 
                                       \Big[  -\frac{\nabla ^2}{2m_{\rm{el}}}  -(JS + \frac{B}{2} )\sigma^z   \Big] c({\mathbf{x}},t). 
 \label{eqn:e5}                                      
\end{eqnarray}                                      
The variable $m_{\rm{el}}$ denotes the effective mass of a conduction electron.
Eq. (\ref{eqn:e5}) shows that
\begin{eqnarray}
JS 
\end{eqnarray}
acts as an effective magnetic field.

The dynamics at the interface  is described by 
the Hamiltonian  ${\cal{H}}_{\rm{ex}} ^{\prime}    $ ;
eq. (\ref{eqn:e4})  shows that localized spins at the interface lose  spin angular momentum by emitting  a magnon
and  a conduction electron flips from down to up by absorbing the spin angular momentum 
(see Fig. \ref{fig:pumping}), and vice versa.
This Hamiltonian ${\cal{H}}_{\rm{ex}} ^{\prime}    $, which describes the interchange of spin angular momentum
between localized spins and conduction electrons,     
is essential  to spin pumping.
 Therefore we clarify the contribution of magnons accompanying this exchange interaction to spin pumping.  
This is the main purpose of this paper.
Here  it should be noted that
 we treat localized spins as not classical variables \cite{xiao}  but magnon degrees of freedom.
As the result, we can microscopically capture the (non-equilibrium) spin-flip dynamics
on the basis of  the rigorous quantum mechanical theory.

This paper is structured as follows.
First, through the  Heisenberg equation of motion,  the thermal spin transfer torque
which breaks the spin conservation law for conduction electrons 
is defined in  sec. \ref{sec:torque}.
Second, we  evaluate it through the Schwinger-Keldysh formalism 
at finite temperature in  sec. \ref{sec:kel}.
Last  
we discuss why thermal spin pumping does not cost any applied magnetic fields
 in   sec. \ref{sec:pumping}, with pointing out the distinction 
from the farseeing  work by Adachi et al. \cite{adachi}.

\section{Thermal spin transfer torque}
\label{sec:torque}

\subsection{Definition}
\label{subsec:definition}

The thermal spin transfer torque (TSTT)  \cite{adachi,TSTT,TSTT2},  ${\mathcal{T}}_{\rm{s}}^z$, is defined as the term 
which  breaks the spin conservation law for conduction electrons;
 \begin{eqnarray}
 \dot \rho_{\rm{s}}^z  +   \nabla \cdot {\mathbf{j}}_{\rm{s}}^{z} =  {\mathcal{T}}_{\rm{s}}^z ,
\label{eqn:6}   
 \end{eqnarray}
 where the dot denotes the time derivative,  ${\mathbf{j}}_{\rm{s}}$  is the spin current density\cite{takeuchi}, and
$\rho_{\rm{s}}^z$ represents the z-component of the spin density.
  We here have defined  the spin density of the system   as the expectation value (estimated for the total Hamiltonian, {$\cal{H}$});
  \begin{eqnarray}
  \rho_{\rm{s}}^z \equiv  \langle  c^{\dagger }  \sigma ^z c/2  \rangle.
  \end{eqnarray}
In this paper, we focus on the z-component of the TSTT.

Through the Heisenberg equation of motion,
the  z-component of the TSTT   is defined as 
\begin{eqnarray}
{\mathcal{T}}_{\rm{s}}^z  = iJa_0^3  \sqrt{\frac{\tilde S}{2}} \langle  a^{\dagger }({\mathbf{x}},t) 
                             c^{\dagger }  ({\mathbf{x}},t) \sigma^{+}   c({\mathbf{x}},t )    
                          -  a({\mathbf{x}},t)  c^{\dagger } ({\mathbf{x}},t)  \sigma^{-}   c ({\mathbf{x}},t)\rangle.
\label{eqn:e7}                         
\end{eqnarray}
This term arises from $   { \cal{H}}_{\rm{ex}}^{\prime} $,
which consist of electron spin-flip  operators;   
\begin{eqnarray}
 {\mathcal{T}}_{\rm{s}}^z  =   [\rho_{\rm{s}}^z,  { \cal{H}}_{\rm{ex}}^{\prime} ]/i.
 \end{eqnarray}
 Thus, eq. (\ref{eqn:6}) shows that
the TSTT   (${\mathcal{T}}_{\rm{s}}^z >0$)  can be understood as the number density of conduction electrons
which flip from down to up  per a unit of time \cite{zubarev}, and vice versa.
In addition,
the TSTT operates the coherent magnon state  \cite{coherent2}.

\subsection{Pumped net spin current}
\label{subsec:pumped}

In this subsection,
we clarify the relation between the TSTT and the pumped net spin current.
As discussed in the last subsection,
the spin conservation law for conduction electrons is broken
due to the interaction ${\cal{H}}_{\rm{ex}}^{\prime}$; 
\begin{eqnarray}
\dot \rho_{\rm{s}}^z  +   \nabla \cdot {\mathbf{j}}_{\rm{s}}^{z} =  {\mathcal{T}}_{\rm{s}}^z.
\label{eqn:e51}  
\end{eqnarray}
Thus one cannot simply view the time derivative of the spin density for conduction electrons, $ \dot \rho_{\rm{s}}^z$,
as the spin current density.

In respect to Planck's constant (we here partially recover $\hbar $),
the time derivative of the spin density and the TSTT satisfy the relation \cite{nambu,coleman};
\begin{eqnarray}
\frac{\dot \rho_{\rm{s}}^z}{{\mathcal{T}}_{\rm{s}}^z} = {\cal{O}}(\hbar ).
\end{eqnarray}
Therefore
$\dot \rho_{\rm{s}}^z$ is negligible in comparison with ${\mathcal{T}}_{\rm{s}}^z$  at the semiclassical regime, where our interest lies.
As the result,
the spin continuity equation, eq. (\ref{eqn:e51}), becomes
 \begin{eqnarray}
 {\mathcal{T}}_{\rm{s}}^z  = \nabla \cdot {\mathbf{j}}_{\rm{s}}^{z}.
\label{eqn:e52}  
\end{eqnarray}
Then by integrating over the interface,
we can evaluate the pumped net spin current, $\int  {\mathbf{j}}_{\rm{s}}^{z}  \cdot  d{{\mathbf{S}}_{\rm{interface}}}$;
 \begin{eqnarray}
      \int_{{\mathbf{x}}\in \rm{(interface)}}    d{\mathbf{x}}   \  {\mathcal{T}}_{\rm{s}}^z 
&=\int_{{\mathbf{x}}\in \rm{(interface)}}    d{\mathbf{x}}   \    \nabla \cdot {\mathbf{j}}_{\rm{s}}^{z}  \\
&= \int  {\mathbf{j}}_{\rm{s}}^{z}  \cdot  d{{\mathbf{S}}_{\rm{interface}}}.
\label{eqn:e53}  
\end{eqnarray}

In addition, conduction electrons  cannot  enter the ferromagnet, which is an insulator  \cite{spinwave}. 
Thus the net spin current  pumped into the non-magnetic metal can be calculated
by integrating the TSTT  over  the interface, eq. (\ref{eqn:e53}).

From now on, we focus on  $ {\mathcal{T}}_{\rm{s}}^z$  and 
qualitatively  clarify the behavior of the thermal spin pumping effect mediated by magnons,  
at  room temperature in the semiclassical regime, in sections \ref{sec:kel} and \ref{sec:pumping}.

\subsubsection{The spin continuity equation for the whole system}

It will be useful to point out that 
the spin conservation law for localized spins (i.e. magnons) is also broken.
The magnon continuity equation for localized spins  \cite{nakata} reads 
\begin{eqnarray}
\dot \rho_{\rm{m}}^z  +   \nabla \cdot {\mathbf{j}}_{\rm{m}}^{z} =  {\mathcal{T}}_{\rm{m}}^z ,
\label{eqn:e54}  
\end{eqnarray}
 where ${\mathbf{j}}_{\rm{m}}$  is the magnon current density, and
$\rho_{\rm{m}}^z$ represents the z-component of the magnon density.
 We  have  defined  the magnon density of the system also  as the expectation value (estimated for the total Hamiltonian, {$\cal{H}$});
 \begin{eqnarray}
 \rho_{\rm{m}}^z \equiv  \langle  a^{\dagger } a  \rangle.
 \end{eqnarray}
 In addition, we call ${\mathcal{T}}_{\rm{m}}^z$ the magnon source term  \cite{nakata},
which breaks the magnon conservation law.
This term arises  also from ${\cal{H}}_{\rm{ex}}^{\prime}$;
 \begin{eqnarray}
 {\mathcal{T}}_{\rm{m}}^z = [\rho_{\rm{m}}^z, {\cal{H}}_{\rm{ex}}^{\prime}]/i.
 \end{eqnarray}

Through the Heisenberg equation of motion,
the magnon source term can be determined 
and it satisfies the relation;
\begin{eqnarray}
 {\mathcal{T}}_{\rm{m}}^z = {\mathcal{T}}_{\rm{s}}^z.
 \label{eqn:e61}  
 \end{eqnarray}
Then the z-component of the spin continuity equation for the total system (i.e. conduction electrons and magnons) becomes
\begin{eqnarray}
\dot {\rho} _{\rm{total}}^z + \nabla \cdot  {\mathbf{j}}_{\rm{total}}^z =0,
\label{eqn:e55}  
\end{eqnarray}
where the density of the total spin angular momentum, $ {\rho} _{\rm{total}}^z$, is defined as
\begin{eqnarray}
{\rho} _{\rm{total}}^z \equiv  {\rho} _{\rm{s}}^z -  {\rho} _{\rm{m}}^z,
\end{eqnarray}
and consequently
the z-component of the total spin current density,  ${\mathbf{j}}_{\rm{total}}^z$, becomes
\begin{eqnarray}
 {\mathbf{j}}_{\rm{total}}^z  =    {\mathbf{j}}_{\rm{s}}^z -  {\mathbf{j}}_{\rm{m}}^z
 \end{eqnarray}
 (note that, $S^z = \tilde S - a^{\dagger } a$,  via the Holstein-Primakoff transformation  in sec. \ref{sec:intro}).
The spin continuity equation for the whole system, eq. (\ref{eqn:e55}),
 means that 
though each spin conservation law for electrons and magnons is broken (see eqs. (\ref{eqn:e51}) and (\ref{eqn:e54})),
the total spin angular momentum  is, of course, conserved \cite{zubarev}.

\subsubsection{The work by Bender et al.}

Last, let us mention a recent preprint \cite{bender} by Bender et al.,
where the authors consider a similar problem.
We have chosen a different definition of the pumped spin current, for
reasons now explained.

Though they have simply recognized the time derivative of the spin density for localized spins,
\begin{eqnarray}
\dot \rho_{\rm{m}}^z,
\end{eqnarray}
as the spin current,\footnote{ 
Note that we have adopted our notation.
The variable $dS_{L}^z/dt$ in Ref. \cite{bender} corresponds to $\dot \rho_{\rm{m}}^z$.}
it reads
\begin{eqnarray}
\dot \rho_{\rm{m}}^z   
{\stackrel{{\rm{eq}}. (\ref{eqn:e55})}{=}}  
\dot \rho_{\rm{s}}^z +\nabla \cdot ({\mathbf{j}}_{\rm{s}}^z -  {\mathbf{j}}_{\rm{m}}^z).  
\label{eqn:e77}  
\end{eqnarray}
Thus it is clear that
even when the total spin angular momentum is conserved (eq. (\ref{eqn:e55})),
$\dot \rho_{\rm{m}}^z$   is not directly related to the spin current itself, ${\mathbf{j}}_{\rm{s}}^z$.
That is,
$\dot \rho_{\rm{m}}^z$ includes other contributions arising from  $\dot \rho_{\rm{s}}^z$ and ${\mathbf{j}}_{\rm{m}}^z$
as well as ${\mathbf{j}}_{\rm{s}}^z$.
Therefore
the definition  of the pumped spin current  by Bender et al. \cite{bender} is, in any regime, inadequate to their and our case; 
the mixture of the Bose (magnon) gas and Fermi (conduction electron) one.

That is why,
we have adopted different definition of the pumped spin current, eq. (\ref{eqn:e53}),
and evaluate the TSTT.

\section{Schwinger-Keldysh formalism}
\label{sec:kel}

The interface  is,  in general,   a weak coupling regime   \cite{AndoPumping};
the exchange interaction, $J$, is supposed to be
 smaller than the Fermi energy and
the exchange interaction among ferromagnets.
Thus  $ { \cal{H}}_{\rm{ex}}^{\prime} $ can be treated as a perturbative term.

Through the standard procedure of the Schwinger-Keldysh (or non-equilibrium) Green's function  \cite{ramer,kamenev,kita},
the Langreth method  \cite{haug,tatara,new},
the TSTT  can be  evaluated as 
\begin{eqnarray}
   {\mathcal{T}}_{\rm{s}}^z &=&   2i J^2 a_0^3 S   \int \frac{{d{\mathbf{k}}_1}}{(2\pi)^3}   \int \frac{{d{\mathbf{k}}_2}}{(2\pi)^3}
         \int    \frac{d\omega _1}{2\pi}  \int \frac{d\omega _2}{2\pi}                                                                                        \nonumber    \\    
&\times &[  {\mathcal{G}}^{\rm{>}}_{\uparrow , {\mathbf{k}}_2, \omega _2}      {\rm{G}}^{\rm{>}}_{{\mathbf{k}}_1, \omega _1}    
   {\mathcal{G}}^{\rm{<}}_{ \downarrow , {\mathbf{k}}_1 + {\mathbf{k}}_2, \omega _1+ \omega _2}     
-   {\mathcal{G}}^{\rm{<}}_{\uparrow , {\mathbf{k}}_2, \omega _2}    {\rm{G}}^{\rm{<}}_{{\mathbf{k}}_1, \omega _1} 
   {\mathcal{G}}^{\rm{>}}_{ \downarrow , {\mathbf{k}}_1 + {\mathbf{k}}_2, \omega _1+ \omega _2}   ]                                                                    
  + {\cal{O}}(J^3).
\label{eqn:e8}
\end{eqnarray}
The variable   $  {\mathcal{G}}^{< (>)} $  
is the fermionic lesser (greater)  Green's function,
and $  {\rm{G}}^{< (>)}$ is the bosonic one.
We here have  taken the extended time defined on the Keldysh contour  \cite{kita,haug,tatara}, c, 
on the forward path ${\rm{c}}_{\rightarrow }$; 
$ c =  {\rm{c}}_{\rightarrow } + {\rm{c}}_{\leftarrow } $.
Even when the time is located  on the backward path ${\rm{c}}_{\leftarrow }$, 
the result of the calculation does not change
because each Green's function is not independent;
$ {\mathcal{G}}^{\rm{r}}  - {\mathcal{G}}^{\rm{a}}  = {\mathcal{G}}^{\rm{>}}  - {\mathcal{G}}^{\rm{<}} $,
where $ {\mathcal{G}}^{\rm{r} (\rm{a})}$ represents the  retarded (advanced) Green's function  \cite{nakata}.
This relation comes into effect also for the bosonic case \cite{kita}.  

Each Green's function reads as follows  \cite{kamenev};
\begin{eqnarray} 
{\rm{G}}^{\rm{<}}_{   {\mathbf{k}}, \omega } & =& -2 \pi i f_{\rm{B}}(\omega) \delta  (\omega  - \omega _{\mathbf{k}}),                \\
{\rm{G}}^{\rm{>}}_{   {\mathbf{k}}, \omega }  &=& -2 \pi i [1+f_{\rm{B}}(\omega)] \delta  (\omega  - \omega _{\mathbf{k}}),               \\
{\mathcal{G}}^{\rm{<}}_{\sigma ,   {\mathbf{k}}, \omega }  &=& 2 \pi i f_{\rm{F}}(\omega) \delta  (\omega  - \omega _{\sigma , \mathbf{k}}),     \\
{\mathcal{G}}^{\rm{>}}_{\sigma ,   {\mathbf{k}}, \omega }  &=& -2 \pi i [1-f_{\rm{F}}(\omega)] \delta  (\omega  - \omega _{\sigma , \mathbf{k}}),
\end{eqnarray}
where the variables
 $f_{\rm_{B}} (\omega )$ and $f_{\rm_{F}} (\omega )$  are 
the Bose distribution function  and the Fermi one.  
The energy dispersion relation reads
 $  \omega _{\mathbf{k}} \equiv  D {k}^2+B $ and 
$  \omega _{\sigma, \mathbf{k}} \equiv  F k^2- ( JS + B/2 )\sigma - \mu $,
where  $ D   \equiv  1/(2m)    $, $  F\equiv    1/(2m_{\rm{el}})  $,  $ \sigma = +1, -1 (= \uparrow, \downarrow )$, 
and $\mu$  denotes  the chemical potential;
$ \mu (T) =  \epsilon _{\rm{F}}  - {(\pi k_{\rm{B} }T)^2}/({12\epsilon _{\rm{F}}}) + {\cal{O}}(T^4)$.  
The variable $ \epsilon _{\rm{F}}$ represents the Fermi energy.

Consequently, eq.(\ref{eqn:e8}) can be rewritten as 
\begin{eqnarray}
  {\mathcal{T}}_{\rm{s}}^z &=&               { 4{\pi} J^2 a_{0}^3 S }  \int \frac{{d{\mathbf{k}}_1}}{(2\pi)^3} 
                                                                      \int \frac{{d{\mathbf{k}}_2}}{(2\pi)^3}  \int d\omega _1  \int d\omega _2                                                             \nonumber     \\                                   
                                               &\times &  \delta (\omega _1- \omega _{{\mathbf{k}}_1})   \delta (\omega _2- \omega _{\uparrow ,{\mathbf{k}}_2}) 
                                                                      \delta (\omega _1 + \omega _2- \omega _{\downarrow , {\mathbf{k}}_1 +{\mathbf{k}}_2})                                         \nonumber     \\
                                             &\times  &    \Big\{  [1+f_{\rm{B}}(\omega _1)] f_{\rm{F}}(\omega _1 + \omega _2)[1-f_{\rm{F}}(\omega _2) ]                                 
                                             -                  f_{\rm{B}}(\omega _1)     f_{\rm{F}}(\omega _2)[1- f_{\rm{F}}(\omega _1 + \omega _1)]                                                           
                                                                      \Big\}                                                                                                                                                                                                     \\
                                            &=&                  { 4{\pi} J^2 a_{0}^3 S }  \int \frac{{d{\mathbf{k}}_1}}{(2\pi)^3}   \int \frac{{d{\mathbf{k}}_2}}{(2\pi)^3}                         
                                                   \delta ( \omega _{{\mathbf{k}}_1} + \omega _{\uparrow ,{\mathbf{k}}_2} 
                                                                      -  \omega _{\downarrow , {\mathbf{k}}_1 +{\mathbf{k}}_2})                                                                                                    \nonumber    \\
                                          &\times &        \Big\{  f_{\rm{F}}( \omega _{{\mathbf{k}}_1} +\omega _{\uparrow ,{\mathbf{k}}_2})
                                                                        [1-f_{\rm{F}}(\omega _{\uparrow ,{\mathbf{k}}_2}) ]                                                                                                                 
                                          +                    f_{\rm{B}}(\omega _{{\mathbf{k}}_1})  f_{\rm{F}}( \omega _{{\mathbf{k}}_1} +\omega _{\uparrow ,{\mathbf{k}}_2})
                                                                      - f_{\rm{B}}(\omega _{{\mathbf{k}}_1})  f_{\rm{F}}(\omega _{\uparrow ,{\mathbf{k}}_2})
                                                                           \Big\}.                       
\label{eqn:e9}
\end{eqnarray}

\

\noindent{\textbf{Spin-flip condition}}

The delta function in eq. (\ref{eqn:e9})
represents the condition for spin-flip between conduction electrons and magnons.
The modes  (i.e.  ${\mathbf{k}}_1 $ and ${\mathbf{k}}_2$)   which do not satisfy this condition cannot contribute to thermal spin pumping.

The delta function reads
\begin{eqnarray}
\delta ( \omega _{{\mathbf{k}}_1} + \omega _{\uparrow ,{\mathbf{k}}_2}- \omega _{\downarrow , {\mathbf{k}}_1 +{\mathbf{k}}_2})      
 & =& \delta \Big( (D-F)k_1^2 -2F {\mathbf{k}}_1 \cdot  {\mathbf{k}}_2 -2JS \Big)          \label{eqn:e10-3}     \\
 &=& \frac{1}{2Fk_1 k_2} \delta  \Big(   {\rm{cos}}\theta   - \frac{(D-F)k_1^2-2JS}{2Fk_1k_2}   
                                                     \Big),   \label{eqn:e10-2}      
 \end{eqnarray}
 where
 $ {\rm{cos}}\theta      \equiv   {{\mathbf{k}}_1\cdot {\mathbf{k}}_2}/(k_1k_2)    $.      
 Eq. (\ref{eqn:e10-2}) holds true on the condition; $k_1\not=0, k_2 \not=0$, and $F\not=0 $.
This condition can be justified because
the zero-mode for conduction electrons ($k_2=0$) originally cannot contribute to spin pumping which is the  low energy dynamics;
in order to excite the zero-mode so as to become relevant to spin pumping,
it costs vast energy which amounts to the Fermi energy.  
Such a (relatively high energy)  dynamics is out of  the system we focus on, ${\cal{H}}$.  
In addition, when the zero mode for magnons  ($k_1 =0$)  is substituted into  eq. (\ref{eqn:e10-3}),
it gives zero because of the finite effective magnetic fields $JS (\not=0)$.
Thus the zero-mode of magnons also  originally  cannot contribute to spin pumping and are eliminated. 
Then we  are allowed to calculate eq. (\ref{eqn:e9}) on the condition; $k_1\not=0$ and $k_2 \not=0$.

Consequently by using eq. (\ref{eqn:e10-2}),
the TSTT (eq. (\ref{eqn:e9})) can be rewritten as 
\begin{eqnarray}
 \frac{ 4 \pi^3 DF^2}{   a_0^3 S  \epsilon _{\rm{F}} ^4  }  {\mathcal{T}}_{\rm{s}}^z    
&=&  \int_{\searrow 0}^{\infty }d\bar{k}_1 \int_{\searrow 0}^\infty d\bar{k}_2  \bar {\mathcal{T}}_{\rm{s}}^z ({\bar{k}}_1, {\bar{k}}_2)  \\
&\equiv &  \int_{\searrow  0}^{\infty }d\bar{k}_{1}   \bar {\mathcal{T}}_{\rm{s}}^z ({\bar{k}}_{1}),     \\
&\equiv &    \bar {\mathcal{T}}_{\rm{s}}^z,
\label{eqn:e12}
\end{eqnarray}
where 
\begin{eqnarray}
  \bar {\mathcal{T}}_{\rm{s}}^z  ({\bar{k}}_1,  {\bar{k}}_2) 
&\equiv &  {\bar{J}}^2 
 \int_{-1}^{1} d\zeta  \ \delta \Big( \zeta  - \frac{ (1-\frac{F}{D}){\bar k}_1^2  -2\bar{J} S   }{2\sqrt{\frac{F}{D}}{\bar k}_1 {\bar k}_2 }    \Big) 
 \cdot  \bar{k}_1 \bar {k}_2                                                         \nonumber      \\
&\times & \Bigg\{ - \frac{ 1}{{\rm{e}}^{(\bar{k}_1^2+\bar B)/{\bar {T}}_{\rm{m}}}-1}  
 \cdot   \frac{1}{{\rm{e}}^{(\bar{k}_2^2-\bar {J} S-{\bar B}/2 -1+\pi^2 {\bar {T}_{\rm{s}}}^2/12)/{\bar {T}}_{\rm{s}}}+1}          \nonumber           \\
&+& \Big[  1 - \frac{1}{{\rm{e}}^{(\bar{k}_2^2-\bar {J} S-{\bar B}/2 -1+\pi^2 {{\bar {T}}_{\rm{s}}}^2/12)/{{\bar {T}}_{\rm{s}}}}+1}  
            +  \frac{1}{{\rm{e}}^{(\bar{k}_1^2+\bar B)/{\bar {T}}_{\rm{m}}}-1}    \Big]                                                                            \nonumber           \\
 &\times &   \frac{  1  }{{\rm{e}}^{( \bar{k}_1^2 +  \bar{k}_2^2-\bar {J} S +{\bar B}/2 -1+\pi^2 {{\bar {T}_{\rm{s}}}^2/12)/{\bar {T}}_{\rm{s}}}}+1}\Bigg\}.      
\label{eqn:e13}                   
\end{eqnarray}
We here  have defined a variable, $ \zeta  \equiv {\rm{cos}} \theta  $, 
and have introduced dimensionless variables;
$ \bar{k}_1 \equiv \sqrt{ D/ \epsilon _{\rm{F}} } k_1 ,   
\bar{k}_2 \equiv \sqrt{ F/ \epsilon _{\rm{F}} } k_2,
\bar B \equiv  B/{\epsilon _{\rm{F}}}, 
   \   \bar {J} \equiv   {J}/{\epsilon _{\rm{F}}},
{\bar T}_{\rm{m(s)}} \equiv T_{\rm{m(s)}}/T_{\rm{F}}\equiv k_{\rm{B}}T_{\rm{m(s)}}/{\epsilon _{\rm{F}}}$,
where
$k_{\rm{B}}$ denotes the Boltzmann constant.
The variable $ T_{\rm{m(s)}}$ is the effective local temperature of magnons (conduction electrons)  \cite{uchidathermal,adachi,xiao},
and 
\begin{eqnarray}
\bar {\mathcal{T}}_{\rm{s}}^z  ({\bar{k}}_1, {\bar{k}}_2) 
\end{eqnarray}
 represents the dimensionless TSTT
in the wavenumber space for magnons and conduction electrons;
\begin{eqnarray}
\bar {\mathcal{T}}_{\rm{s}}^z ({\bar{k}}_{1})
\end{eqnarray}
 denotes the dimensionless TSTT
in the wavenumber space for magnons, ${\bar{k}}_{1}$,
after integrating over the wavenumber space for conduction electrons, ${\bar{k}}_{2}$.
Both quantities, $\bar {\mathcal{T}}_{\rm{s}}^z  ({\bar{k}}_1, {\bar{k}}_2) $ and $\bar {\mathcal{T}}_{\rm{s}}^z ({\bar{k}}_{1})$,
describe the exchange interaction ($J$) and the temperature ($T_{\rm{m(s)}}$)  dependence of the TSTT.

We   set each parameter,   as a typical case,  as follows  \cite{xiao,spinwave, kittel};
$ \epsilon _{\rm{F}} = 5.6  $ eV,
$ B/{\epsilon _{\rm{F}}} = 0 $,
$ F=4$ eV {\AA}$^2$,
$ D=0.3$ eV {\AA}$^2$,
$S=1/2$.
Here it should be noted that 
we do not apply magnetic fields along the quantization axis; 
\begin{eqnarray}
B=0.
\end{eqnarray}

\begin{figure}[h]
\begin{center}
\includegraphics[width=8cm,clip]{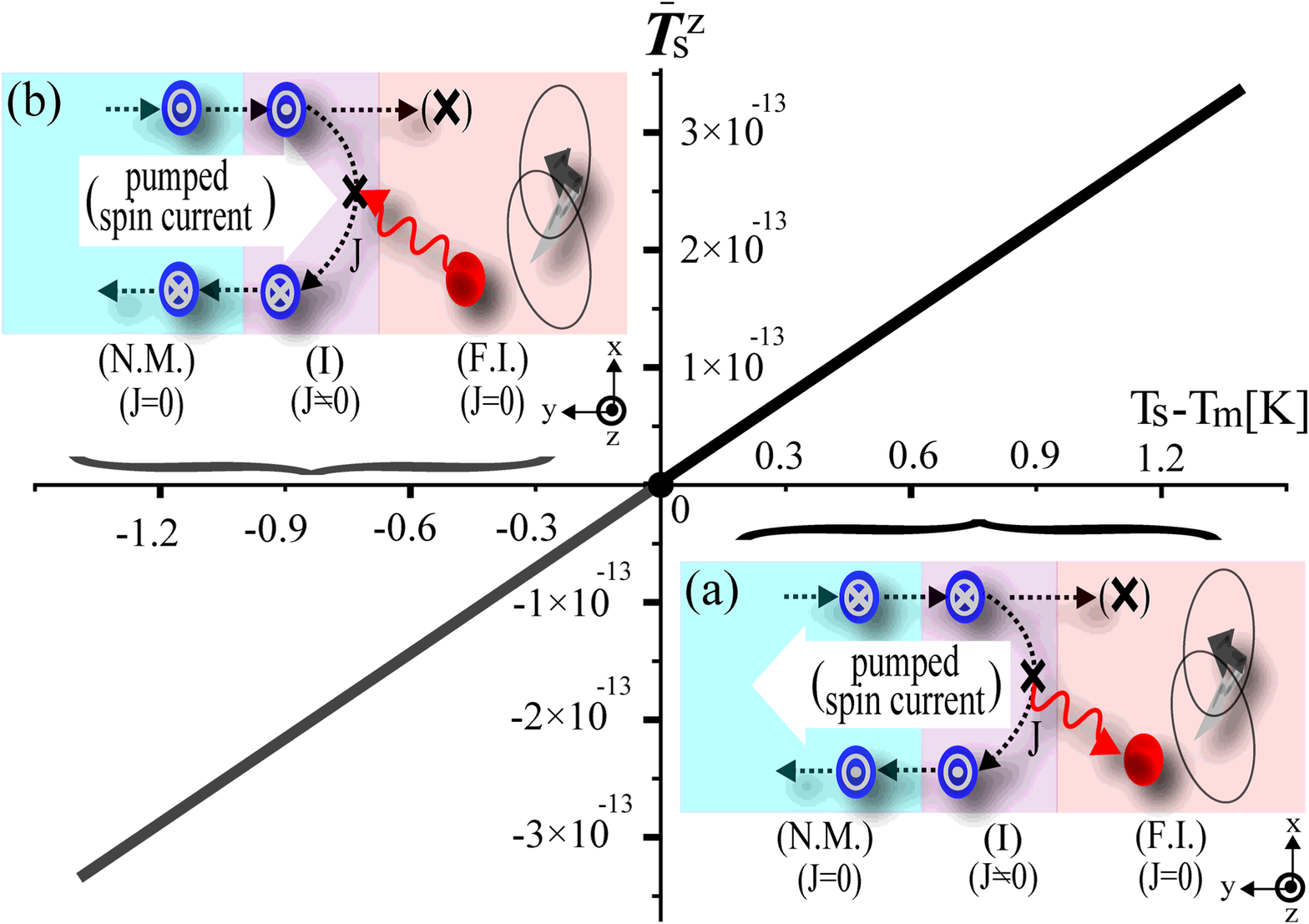}
\caption{
The temperature difference dependence of 
the dimensionless TSTT, $ \bar {\mathcal{T}}_{\rm{s}}^z$,
and the corresponding schematic pictures.
Each parameter reads 
${\bar {J}} = 0.002$ and
$ T_{\rm{s}} = 300  $ K.
When the effective temperature of magnons is lower  than that of conduction electrons,
localized spins at the interface lose  spin angular momentum by emitting  magnons 
and conduction electrons flip from down to up by absorbing the momentum (a),
and vice versa (b). 
 \label{fig:thermal} }
\end{center}
\end{figure}

\begin{figure}[h]
\begin{center}
\includegraphics[width=10cm,clip]{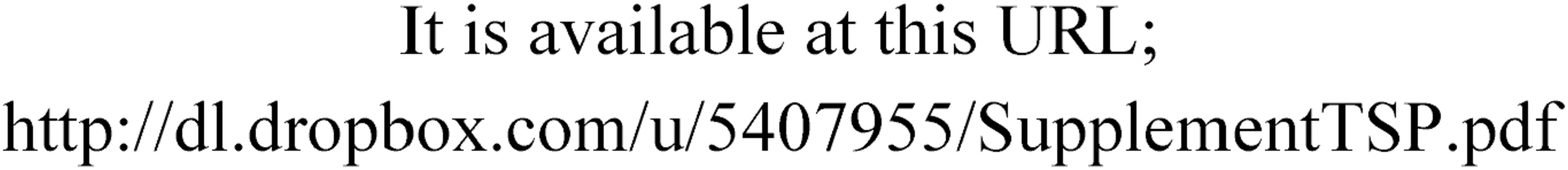}
\caption{
(a) 
The spin-flip condition via magnons;
$ z ({\bar k}_1, {\bar k}_2)=  \zeta ^{\prime} \equiv  [{ (1-{F}/{D}){\bar k}_1^2  -2\bar{J} S   }]({2\sqrt{{F}/{D}}{\bar k}_1 {\bar k}_2 })^{-1}$,
where ${\bar {J}} = 0.002$.
Magnons at (near) the zero-mode cannot contribute to thermal spin pumping
because they do not satisfy the  spin-flip condition, eq. (\ref{eqn:e10-3}).
(b) 
The TSTT in the wavenumber space for 
conduction electrons and magnons, $ \bar {\mathcal{T}}_{\rm{s}}^z  ({\bar{k}}_1,  {\bar{k}}_2) $. 
Each parameter reads 
${\bar {J}} = 0.002$,
$ T_{\rm{s}} = 300  $ K, and
$ T_{\rm{s}}-  T_{\rm{m}} = 1.2  $ K.
A sharp peak exists on the Fermi wavenumber.
(c)
The TSTT in the wavenumber space for magnons, $ \bar {\mathcal{T}}_{\rm{s}}^z ({\bar{k}}_{1})$;
the condition is the same with (b).
The higher  the effective magnon temperature becomes,
the longer wavenumber of magnons becomes relevant to thermal spin pumping.
 \label{fig:zero} }
\end{center}
\end{figure}

\section{Thermal spin pumping effect}
\label{sec:pumping}

Fig. \ref{fig:thermal} shows that
under the thermal equilibrium condition  where  temperature difference  does not exist  between ferromagnet and non-magnetic metal,
spin currents cannot be pumped because of the balance between thermal fluctuations in ferromagnet and those in non-magnetic metal  \cite{uchidathermal,adachi,xiao}.
In addition, it can be concluded  that 
the pumped spin current
is proportional to the temperature difference between the magnon and conduction electron temperatures (i.e. $T_{\rm{s}}-T_{\rm{m}}$);
when the effective temperature of magnons is lower  than that of conduction electrons (see Fig. \ref{fig:thermal} (a)),
localized spins  at the interface  lose  spin angular momentum by emitting  magnons 
and conduction electrons flip from down to up by absorbing  all  the  emitted momentum  \cite{zubarev},
and vice versa (see Fig. \ref{fig:thermal} (b)).
This result exhibits the good agreement with the work by Xiao et al.  \cite{xiao};
they have reached this result by combining the spin pumping theory proposed by  Tserkovnyak et al. \cite{mod2}
with  the Landau-Lifshitz-Gilbert equation.

Figs. \ref{fig:zero} (a) and (c) show that 
magnons at (near) the zero-mode cannot contribute to thermal spin pumping
because they do not satisfy the  spin-flip condition between conduction electrons and magnons,
due to the finite effective magnetic field $JS$.
 (see eqs. (\ref{eqn:e10-3}),  (\ref{eqn:e13}), and Fig. \ref{fig:zero} (a) ).

\

\noindent{\textbf{The distinction from the work 
by Xiao et al. and Adachi et al.}}

Let us mention that  we have set $B=0$.
That is,
 a spin current can be generated via the thermal spin pumping effect without any applied magnetic fields.
{ This point cannot  be obtained by Xiao et al. \cite{xiao}.   }
The pumped spin current is proportional to the temperature difference between the magnon and conduction electron temperatures;
inhomogeneous thermal fluctuations   induce a net spin current  \cite{uchidathermal}. 
This is the main difference from the quantum spin pumping  effect   \cite{QSP}.

Last we should discuss the  distinction from the important work by Adachi et al.  \cite{adachi}, 
with emphasizing that
they have already studied thermal spin pumping via magnons before our study.
They have pointed out that
the approach   by using  the stochastic Landau-Lifshitz-Gilbert equation coupled with the Bloch equation
is equivalent to the one by the Schwinger-Keldysh formalism (i.e. linear-response theory) 
in the classical regime where  quantum fluctuations are negligible.
This fact has already been confirmed also by the numerical calculation  \cite{ohe}.
Though they have studied the thermal spin pumping effect mediated by magnons via the Schwinger-Keldysh formalism
by the same procedure with our work,
unfortunately we have doubts the validity of their calculation;
with reflecting the statistical properties,
the Keldysh Green's function  \cite{kita}  for fermions (${\cal{G}}^{\rm{K}}$) should  be\cite{kamenev}
\begin{eqnarray}
{\cal{G}}^{\rm{K}}_{   {\mathbf{k}}, \omega } (\equiv   
{\cal{G}}^{\rm{<}}_{   {\mathbf{k}}, \omega }   + {\cal{G}}^{\rm{>}}_{   {\mathbf{k}}, \omega })  
 = 2   i    \ {\rm{Im}}    {\cal{G}}^{\rm{r}}_{   {\mathbf{k}}, \omega }   {\rm{tanh}}(\beta \omega /2),
\end{eqnarray}
 not  $2   i \  {\rm{Im}}   {\cal{G}}^{\rm{r}}_{   {\mathbf{k}}, \omega }   {\rm{coth}}(\beta \omega /2)  $  \cite{adachi}.
The variable $ {\cal{G}}^{\rm{r}}_{   {\mathbf{k}}, \omega }$ denotes the fermionic retarded Green's function
and $\beta $ is  defined as  $\beta \equiv  1/(k_{\rm{B}} T)$.
That is, the fermionic Keldysh Green's function is different from the bosonic one. 
In addition, we would like to mention that
though they have taken a classical approximation,
we have discussed the thermal spin pumping effect
in the semiclassical regime.
Moreover,
we  stress that  thermal spin pumping does not cost any applied magnetic field,
magnetic fields along the quantization axis nor transverse magnetic fields,
because magnons at the zero-mode  are eliminated because of the spin-flip condition, eq. (\ref{eqn:e10-3}).

\section{Summary and discussion}
\label{sec:sum}

We have qualitatively studied thermal spin pumping mediated by magnons in the semiclassical regime.
Pumped spin currents are proportional to the  temperature difference 
between conduction electrons and magnons.
That is,  inhomogeneous    thermal fluctuations induce a net spin current;
when the effective temperature of magnons is lower than that of conduction electrons,
localized spins lose spin angular momentum by emitting magnons and 
conduction electrons flip from down to up by absorbing the momentum, and vice versa.
Thermal spin pumping has the advantage that 
it does not cost any kinds of applied magnetic fields because
magnons at the zero mode are eliminated due to the  spin-flip condition.
This fact will be useful for potential applications
to green information and communication technologies;
spin currents can avoid Joule heating.

Though the behavior of the thermal spin pumping effect mediated by magnons can be
qualitatively captured by calculating the TSTT, 
we recognize that  the theoretical estimation for  the width of the interface,
so called proximity effects,
is essential for the quantitative understanding.
In addition, we are also interested in the contribution of
phonons and that of magnons under a spatially nonuniform magnetization
to spin pumping.

\section{Acknowledgements}
We would like to thank  
K. Totsuka for stimulating the study and useful comments.
We are also grateful to T. Takahashi and Y. Korai  for fruitful discussion.
We are  supported by the Grant-in-Aid for the Global COE Program
"The Next Generation of Physics, Spun from Universality and Emergence"
from the Ministry of Education, Culture, Sports, Science and Technology (MEXT) of Japan.

\

------------------------------------------------------------------------------------------

{\textit{\textbf{Supplement is available at this URL;}}}

{\textbf{http://dl.dropbox.com/u/5407955/SupplementTSP.pdf}}

------------------------------------------------------------------------------------------

\bibliographystyle{unsrt}
\bibliography{PumpingRef}

\end{document}